\documentclass[10pt,aps,twocolumn,prl,amsmath,amssymb,showpacs,superscriptaddress]{revtex4-1}
\usepackage{graphicx}
\pdfoutput=1

\begin{document}
\title{Thin-Film Magnetization Dynamics on the Surface of a Topological Insulator}
\author{Yaroslav Tserkovnyak}
\affiliation{Department of Physics and Astronomy, University of California, Los Angeles, California 90095, USA}
\author{Daniel Loss}
\affiliation{Department of Physics, University of Basel, Klingelbergstrasse 82, CH-4056 Basel, Switzerland}
 
\begin{abstract}
We theoretically study the magnetization dynamics of a thin ferromagnetic film exchange-coupled with a surface of a strong three-dimensional topological insulator. We focus on the role of electronic zero modes imprinted by domain walls (DW's) or other topological textures in the magnetic film. Thermodynamically reciprocal hydrodynamic equations of motion are derived for the DW responding to electronic spin torques, on the one hand, and fictitious electromotive forces in the electronic chiral mode fomented by the DW, on the other. An experimental realization illustrating this physics is proposed based on a ferromagnetic strip, which cuts the topological insulator surface into two gapless regions. In the presence of a ferromagnetic DW, a chiral mode transverse to the magnetic strip acts as a dissipative interconnect, which is itself a dynamic object that controls (and, inversely, responds to) the magnetization dynamics.
\end{abstract}



\maketitle

Following theoretical predictions \cite{pankratovSSC87,*fuPRL07to,*moorePRB07,*royPRB09} and experimental realizations \cite{hsiehNAT08,*xiaNATP09} of three-dimensional topological insulators (TI's), vigorous ongoing activities in this burgeoning field are aimed at introducing spontaneous symmetry breaking mechanisms into the system. This could be accomplished by bulk or surface doping to induce magnetism or superconductivity in the parent (essentially free-electron) TI, or by a heterostructure design wherein symmetry breaking is instilled at the TI surface by a quantum proximity effect. We are following the latter route, considering an insulating ferromagnetic layer (MI) capping the bulk TI, such that the TI surface states are exchange coupled to the collective magnetic moment of the MI. Previous theoretical investigations of a similar TI/MI heterostructure were concerned with current-induced spin torques experienced by a monodomain MI \cite{garatePRL10}, Gilbert damping by a doped TI \cite{yokoyamaPRB10md}, electric charging of magnetic textures \cite{nomuraPRB10}, and the rectification of charge pumping by a monodomain precession \cite{uedaCM11}, all in case of a well-defined  spatially uniform sign of the time-reversal symmetry breaking gap in the TI. The essential physical ingredient underlying the key ideas in these papers is the axion electrodynamics \cite{wilczekPRL87} associated with the TI \cite{qiPRB08to}, with a quantized magnetoelectric coupling that is odd under time reversal. In this Letter, we are interested in salient features associated with dynamic magnetic textures that imprint a spatially inhomogeneous gap onto the TI surface states, both in regard to its magnitude and sign. The latter, in particular, engenders electronic chiral modes at the magnetic domain boundaries \cite{hasanRMP10}, whose hydrodynamics become intricately coupled with magnetic precession.

According to the spin-charge helicity of the TI electronic states, the spin-transfer torques acting on the MI are locked with the self-consistent electronic charge currents in the TI. These currents, in turn, can respond to a combination of electromagnetic fields and fictitious forces induced by MI dynamics, having several distinct contributions: (i) two-dimensional (2D) surface currents related to the half-quantized anomalous Hall effect, whose sign depends on the orientation of the capping magnetic domain, (ii) persistent currents governed by the magnetization texture in the capping MI layer, and (iii) Fermi-level chiral currents along the domain walls (DW's) separating regions with an opposite Hall conductance. As an illustrative example, we will describe how the DW position and an internal coordinate that parametrizes its Bloch-to-N{\'e}el transformation are responding to a chiral TI current flowing along the DW. Considering the inverse charge current pumped by the DW dynamics, we highlight a peculiar structure of the Onsager reciprocity, which reverses the DW magnetization as well as the chirality of the associated electronic mode.

\begin{figure}
\includegraphics[width=\linewidth]{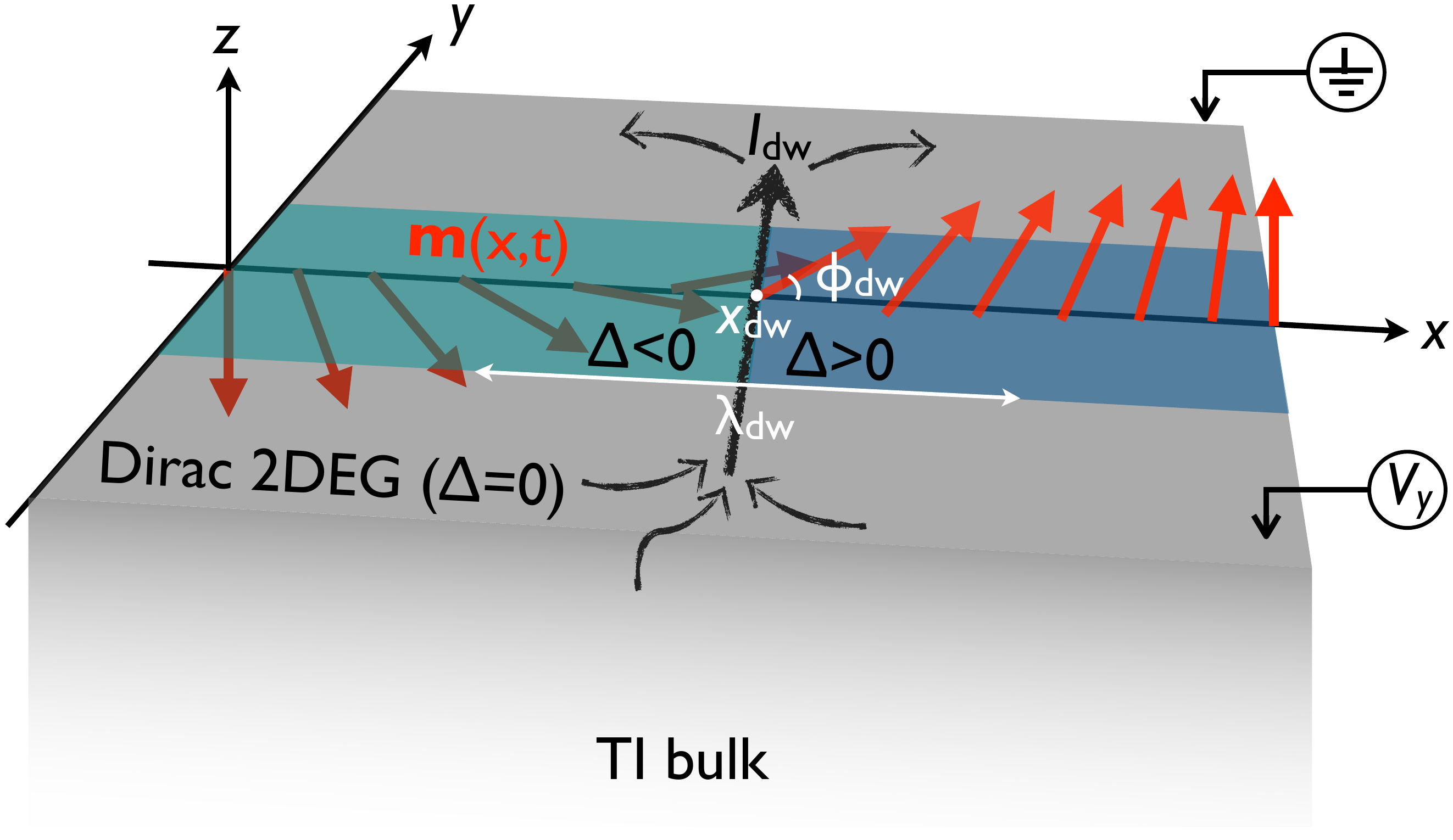}
\caption{Schematic of a domain wall (DW) in a ferromagnetic strip with an out-of-plane easy ($z$) axis anisotropy, deposited on the surface of a topological insulator (TI). The DW (of width $\lambda_{\rm dw}$) is parametrized, according to Eq.~\eqref{pa}, by two \textit{soft} dynamic coordinates: its position $x_{\rm dw}(t)$ and azimuthal angle $\phi_{\rm dw}(t)$. At the DW position, $x_{\rm dw}$, the magnetization $\mathbf{m}$ lies fully in the $xy$ plane (forming angle $\phi_{\rm dw}$ with the $x$ axis). A chiral electron mode (of width $\xi\ll\lambda_{\rm dw}$) formed in the TI under the DW carries transport current $I_{\rm dw}$ at its \textit{exit} point, which is governed by the voltage $V_y$ applied to the TI surface at its \textit{entrance} and the fictitious electromotive force generated by the DW dynamics along its length. An Onsager-reciprocal spin torque affects DW dynamics in the presence of $I_{\rm dw}$.}
\label{fig1}
\end{figure}

Our focus will be centered on a ferromagnetic DW separating regions with out-of-plane magnetization direction deep into the respective domains (which is true for sufficiently thin films, e.g., CoFeB alloys \cite{ikedaNATM10}). See Fig.~\ref{fig1} for a schematic. Let us treat the DW as a stiff solitonic quasi-1D object, parallel to the $y$ axis, whose translational motion and soft internal dynamics can be described by generalized coordinates \cite{thielePRL73,*tretiakovPRL08}. To be specific, we start with the following generic free energy for an isolated magnetic film with magnetic spin texture $\mathbf{m}(\mathbf{r})$ ($|\mathbf{m}|\equiv1$):
\begin{equation}
F_0[\mathbf{m}]=\frac{1}{2}\int d^2r\left\{A\left[(\partial_x\mathbf{m})^2+(\partial_y\mathbf{m})^2\right]-Km_z^2\right\}\,,
\label{F0}
\end{equation}
where $A$ is the exchange stiffness parameter and $K>0$ is the out-of-plane anisotropy constant. A one-dimensional DW running along the $y$ axis and separating magnetic domains with $m_z=\pm1$ at $x\to\pm\infty$, which minimizes free energy \eqref{F0}, is then given in polar angles by
\begin{equation}
\theta(\mathbf{r})=2\tan^{-1}e^{-(x-x_{\rm dw})/\lambda_{\rm dw}}\,,\,\,\,\phi(\mathbf{r})=\phi_{\rm dw}\,,
\label{pa}
\end{equation}
which parametrize position $\mathbf{r}\equiv(x,y)$-dependent magnetization direction $\mathbf{m}=(\sin\theta\cos\phi,\sin\theta\sin\phi,\cos\theta)$. $\lambda_{\rm dw}=\sqrt{A/K}$ is the DW width. The DW energy is degenerate with respect to the position $x_{\rm dw}$ and the azimuthal angle $\phi_{\rm dw}$. In particular, if $\phi=0$ or $\pi$ we have a N{\'e}el wall and if $\phi=\pm\pi/2$ a Bloch wall. In practice, however, the degeneracy with respect to $x_{\rm dw}$ is lifted by spatial pinning fields, while the degeneracy with respect to $\phi_{\rm dw}$ by a homogeneous applied field or spin anisotropy (induced, e.g., by spin-orbit interactions) in the $xy$ plane, which we will take account of below. For sufficiently gentle perturbations of this kind, the zero modes associated with $x_{\rm dw}(t)$ and $\phi_{\rm dw}(t)$ are thus converted into soft collective excitations, which are at the core of our analysis.

We now set out to develop a self-consistent hydrodynamic theory for a DW bound with a gapless chiral mode, which interacts with regions of incompressible TI Hall fluids flanking it on the sides. The introductory material on magnetoelectric properties of the TI, its exchange coupling to the MI, and the emergence of a chiral mode bound to a DW is relegated to the Supplementary Text (ST). From the electronic-structure point of view, the chiral mode patches two quantum Hall regions whose TKNN invariant \cite{thoulessPRL82} changes by unity, between the values of $\pm1/2$ imprinted by the magnetic domains.
The underlying magnetoelectric effect is fundamentally distinct from the one discussed in Ref.~\cite{nomuraPRB10}, as the DW here coexists with the parity-anomaly point $m_z=0$.

A complete picture of our coupled magneto-hydrodynamic system requires us to consider also the spin-transfer torque that is reciprocal to the DW-driven electromotive forces \cite{tserkovPRB09md}. Such torque acting on the magnetization is given, due to the MI/TI exchange (see ST for details)
\begin{equation}
H'=J(m_x\hat{\sigma}_x+m_y\hat{\sigma}_y)+J_\perp m_z\hat{\sigma}_z
\label{Hx}
\end{equation}
by (within the Landau-Lifshitz phenomenology \cite{landauBOOKv9,*gilbertIEEEM04})
\begin{equation}
S\left.\partial_t\mathbf{m}\right|_\tau=\langle\delta_\mathbf{m}H'\rangle\times\mathbf{m}=\left(J\boldsymbol{\sigma}_{xy}+J_\perp\sigma_z\mathbf{z}\right)\times\mathbf{m}\,.
\label{tau}
\end{equation}
Here, $\boldsymbol{\sigma}\equiv(\sigma_x,\sigma_y,\sigma_z)\equiv(\boldsymbol{\sigma}_{xy},\sigma_z)$ is the TI surface spin density (defined by $\boldsymbol{\sigma}=\langle\hat{\boldsymbol{\sigma}}\rangle$, in terms of Pauli matrices $\hat{\boldsymbol{\sigma}}$) and $S$ is the saturation spin density of the ferromagnet. It follows from the Dirac Hamiltonian, $H_0=v(\mathbf{p}-e\mathbf{A})\cdot\mathbf{z}\times\boldsymbol{\hat{\sigma}}+e\varphi$, furthermore, that the in-plane spin density $\boldsymbol{\sigma}_{xy}$ is essentially equivalent to the charge current density, since
\begin{equation}
\mathbf{j}=-\langle\delta_\mathbf{A}H\rangle=ev\mathbf{z}\times\boldsymbol{\sigma}\,\,\,\Rightarrow\,\,\,\boldsymbol{\sigma}_{xy}=(ev)^{-1}\mathbf{j}\times\mathbf{z}\,.
\label{js}
\end{equation}
Eq.~\eqref{js} is an exact identity between the total (i.e., equilibrium plus nonequilibrium) current density and the in-plane spin density of the TI electrons, which is unspoiled by electron-electron interactions and ferromagnetic proximity. In particular, the chiral states, which propagate in the $y$ direction and have spin quantized along the $x$ axis, carry a 2D equilibrium current density that is estimated as (see ST for details)
\begin{equation}
\mathbf{j}\sim(ev/\xi)(\delta_{\rm dw}/2\pi\hbar v)\,\mathbf{z}\times\mathbf{x}\sim(e/2\pi\hbar)J_\perp\partial_xm_z\,\mathbf{y}\,,
\label{jc}
\end{equation}
where we put $\delta_{\rm dw}\sim\xi J_\perp\partial_xm_z$ for the chiral-mode bandwidth in terms of the characteristic (spatial) chiral-mode width $\xi$. From a purely phenomenological perspective, on the other hand, an equilibrium charge current associated with a smooth static texture is given, to the first order in general magnetic inhomogeneities, by
\begin{equation}
\mathbf{j}=\eta J_\perp\mathbf{z}\times\boldsymbol{\nabla}m_z\,,
\label{je}
\end{equation}
which should be valid both near and away from the DW, as long as the current is analytic in the magnetic texture $\mathbf{m}(\mathbf{r})$. This current is time-reversal odd, mirror symmetric (in the $xy$ plane), and divergenceless. We, furthermore, remark that it does not contradict the well-known result for the electromagnetic response of Dirac electrons \cite{redlichPRL84,*jackiwPRD84}, which is exact only for a strictly homogeneous system. The phenomenological coefficient $\eta$ can in general be a function of $(J_\perp m_z)^2$, which we take to be constant in the limit of weak exchange $J_\perp$. Comparing Eqs.~\eqref{jc} and \eqref{je}, we conclude that $\eta\sim e/2\pi\hbar$ in our model, which is suggestive of a universal result (as long as $\xi\ll\lambda_{\rm dw}$).

In the presence of an equilibrium texture-induced current, the spin torque is given by
\begin{equation}
S\left.\partial_t\mathbf{m}\right|_\tau=\delta_\mathbf{m}F_\tau\times\mathbf{m}\,,
\label{tauF}
\end{equation}
in terms of the TI free-energy functional $F_\tau[\mathbf{m}]$ engendered by the MI/TI exchange. According to Eq.~\eqref{je}, this free energy $F_\tau$ can be explicitly found by integrating
\begin{equation}
\delta_{\mathbf{m}_{xy}}F_\tau=J\boldsymbol{\sigma}_{xy}=(J/ev)\mathbf{j}\times\mathbf{z}=(\eta JJ_\perp/ev)\boldsymbol{\nabla}m_z
\label{Fp}
\end{equation}
over $\mathbf{m}_{xy}$, at a fixed $m_z$ \cite{Note2}:
\begin{equation}
F_\tau=(\eta JJ_\perp/ev)\int d^2r\,\mathbf{m}\cdot\boldsymbol{\nabla}m_z+F_\tau'[m_z]\,,
\label{Fz}
\end{equation}
where $F_\tau'$ is a functional of $m_z$ only [which therefore must derive entirely from the $J_\perp$ exchange in Eq.~\eqref{Hx}]. To the leading order in the MI/TI exchange coupling $J_\perp$, $F'_\tau$ contributes merely to the out-of-plane anisotropy $K$ in Eq.~\eqref{F0}, which can be absorbed by a redefinition $K\to K+\mathcal{O}(J_\perp^2)\equiv K_\ast$. Higher-order terms in $F'_\tau$, including those that depend on spatial inhomogeneities in $m_z$, would appear only at order $J_\perp^4$ (while cubic terms are prohibited by the time-reversal invariance). The leading-order MI/TI exchange coupling thus produces an anisotropy $\propto JJ_\perp$, which enhances tendency to form magnetic textures (such as skyrmion lattices), and a texture-independent (easy-axis) out-of-plane anisotropy $\propto J_\perp^2$, corresponding to the first and second terms in Eq.~\eqref{Fz}, respectively.

In addition to the equilibrium current density \eqref{je}, there are also surface currents driven by the real and fictitious electromagnetic fields and the current carried by the gapless chiral mode. The latter may result in dissipation if connected to reservoirs (such as ungapped TI regions). All these currents contribute to the torque \eqref{tau}. [If $I_{\rm dw}$ is the 1D chiral \textit{current}, the corresponding 2D \textit{current density} in the $y$ direction is $j_{\rm dw}\approx I_{\rm dw}\delta(x-x_{\rm dw})$, which is localized on the scale of the chiral-mode width $\xi$.] In particular, as discussed in the ST, the torques arising from the effective electric-field-induced currents correspond to the Chern-Simons action associated with the effective 3-potential $\mathcal{A}_\mu\equiv A_\mu+a_\mu$ [with $A_\mu=(\varphi,-\mathbf{A})$ denoting the physical and $a_\mu=(0,-\mathbf{a})$, where $\mathbf{a}=(J/ev)\mathbf{m}\times\mathbf{z}$, the exchange-induced contributions].

\begin{figure}
\includegraphics[width=\linewidth]{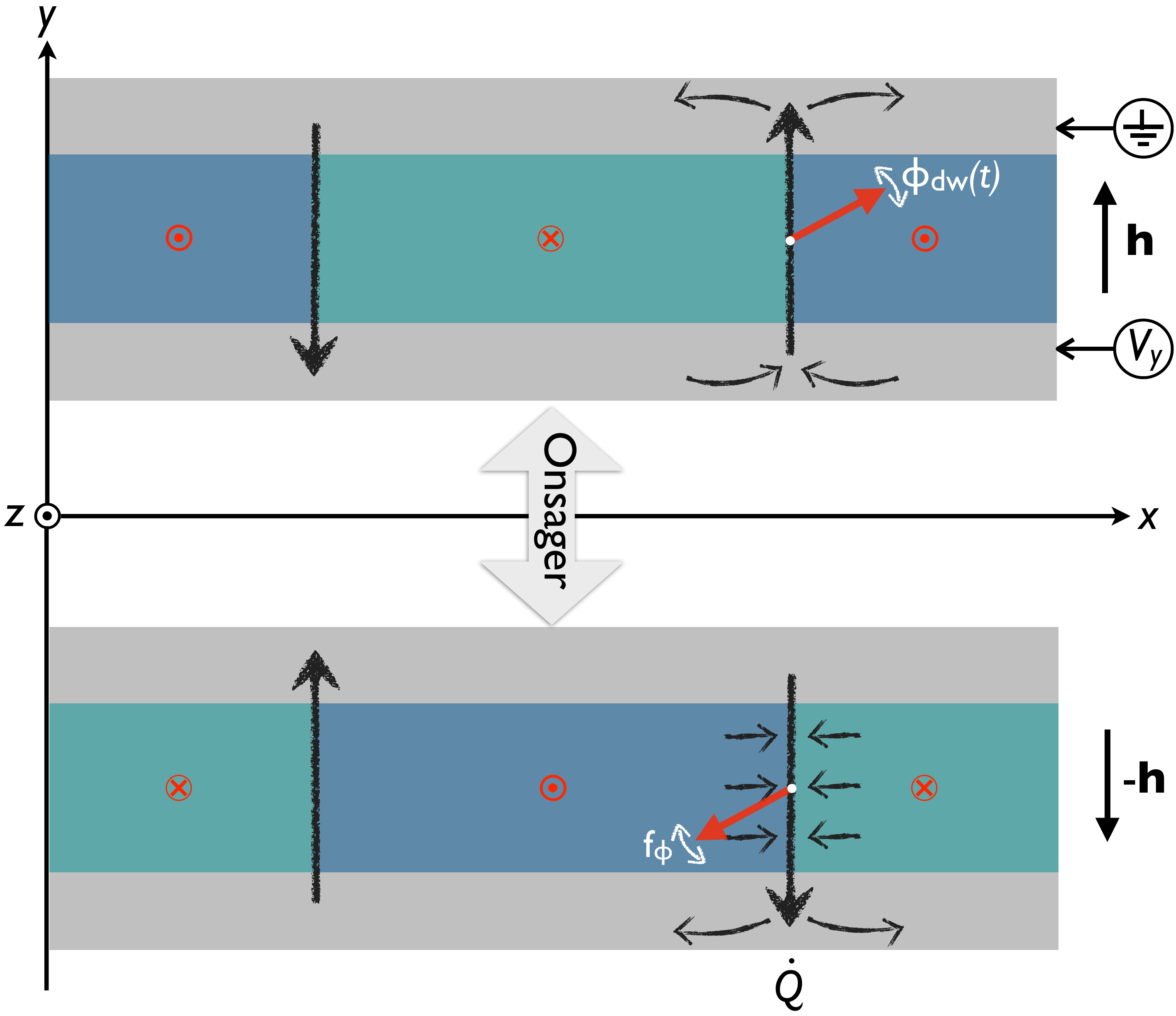}
\caption{The Onsager reciprocity relates voltage-induced DW dynamics (via spin torques) in the top panel [Eq.~\eqref{eom}] to the magnetization-dynamics-generated current (via fictitious electromotive force) in the bottom panel [Eq.~\eqref{Qd}]. Note that the DW's in the bottom panel are mapped back onto their time-reversed parents in the top panel by a $\pi$ rotation in the $xy$ plane. This means that $\dot{Q}$ pumped by $\dot{\phi}_{\rm dw}$ for the right chiral mode is the same in both panels. The left DW is treated as pinned (and thus magnetically inert) in our treatment. However, when the electron-electron interactions are taken into account, electrostatic charge imbalance produced by fictitious forces near one DW could induce currents also along the other DW, making such double-DW system generally coupled.}
\label{fig2}
\end{figure}

Henceforth focusing on the configuration sketched in Fig.~\ref{fig1}, a finite-length DW cuts across a ferromagnetic strip connecting semi-infinite gapless 2D reservoirs flanking its sides. In this case, the reservoirs provide an equilibration and dissipation mechanism for the dc transport. In particular, at low frequencies, the chiral current is given by the Landauer-B{\"u}ttiker formula \cite{nazarovBOOK09}
\begin{equation}
I_{\rm dw}=g_Q\int_{\rm DW}dy\mathcal{E}_y\equiv g_Q\mathcal{V}_y
\label{IwLB}
\end{equation}
for the current in the $y$ direction in response to the total effective field $\mathcal{E}_y$ applied along the DW length (the magnetic strip width). $\mathcal{V}_y=V_y+v_y$ is the corresponding effective voltage ($V_y$ applied and $v_y$ induced by magnetic dynamics) and $g_Q\equiv e^2/h$ is the conductance quantum. The current $I_{\rm dw}$ in Eq.~\eqref{IwLB} is defined at the \textit{exit} point of the chiral mode and, concerning the applied voltage $V_y$, only the effective electric field along the DW \textit{wire} and the chemical potential applied to the \textit{entrance} point of the chiral mode need to be included. The chemical potential at the exit point of the chiral mode, on the other hand, has no effect on the current (at both the exit and the entrance) in the corresponding DW. We emphasize that the current \textit{entering} the chiral mode can generally be distinct from $I_{\rm dw}$. In particular, the dynamically-induced voltage $v_y$ as well as the voltage due to an electric field applied along the DW do not affect the entrance current, which is fully governed by the chemical potential applied at the respective lead. In this case, any imbalance between the currents at the ends of the DW is absorbed by the gapped 2D regions flanking the chiral mode [in accordance with the effective magnetic field $\mathcal{B}_z=B_z+b_z$ \cite{redlichPRL84}, where $b_z\equiv\mathbf{z}\cdot\boldsymbol{\nabla}\times\mathbf{a}=-(J/ev)\boldsymbol{\nabla}\cdot\mathbf{m}$ is the texture-induced field], which we schematically sketch in the bottom panel of Fig.~\ref{fig2}. Since the currents entering and exiting each individual DW thus depend very sensitively on the electrostatic considerations concerning the break-down of the effective electrochemical potentials into the electric and chemical counterparts, we will focus on the noninteracting (i.e., well-screened) electrons driven by a combination of a \textit{chemical}-potential bias at the leads and magnetization dynamics along the DW.

We are now fully equipped to derive the equations of motion for the collective soft DW coordinates $x_{\rm dw}(t)$ and $\phi_{\rm dw}(t)$ that parametrize the DW position and internal structure according to Eq.~\eqref{pa}. In the presence of the (equilibrium and nonequilibrium) current-induced spin torques [corresponding to Eqs.~\eqref{Fz} and \eqref{IwLB}, respectively] as well as a uniform magnetic field $h$ applied in the $y$ direction, the full Landau-Lifshitz-Gilbert (LLG) equation \cite{landauBOOKv9} for the magnetization dynamics becomes
\begin{equation}
S(1+\alpha\mathbf{m}\times)\partial_t\mathbf{m}=\mathbf{m}\times\mathbf{H}_\ast\,,
\label{LLG}
\end{equation}
where $\alpha$ is the dimensionless Gilbert damping constant and the total effective field [including the usual Larmor piece $\mathbf{H}_{\rm eff}\equiv-\delta_\mathbf{m}F_0$ and the spin torques] is given by
\begin{align}
\mathbf{H}_\ast=&A\partial_x^2\mathbf{m}+K_\ast m_z\mathbf{z}+h\mathbf{y}\nonumber\\
&+\eta_\ast\left(\mathbf{z}\partial_xm_x-\mathbf{x}\partial_xm_z\right)-j_\ast\mathbf{x}\delta(x-x_{\rm dw})\,.
\label{Hs}
\end{align}
Here, $\eta_\ast\equiv\eta JJ_\perp/ev$ (with $\eta\sim e/2\pi\hbar$), according to Eq.~\eqref{Fz}, and $j_\ast=(J/ev)\bar{I}_{\rm dw}$, according to Eq.~\eqref{js}. $\bar{I}_{\rm dw}$ is the \textit{average} transport current flowing under the DW along the $y$ axis \cite{Note3}. $\eta_\ast$ and $j_\ast$ thus parametrize the equilibrium and nonequilibrium spin torques, respectively.

The equations of motion for the generalized coordinates $\{q_i\}\equiv \{x_{\rm dw},\phi_{\rm dw}\}$ are derived from Eqs.~\eqref{LLG} and \eqref{Hs} by integrating $\int d^2r\,\partial_{q_i}\mathbf{m}\cdot(\mathbf{m}\times[{\rm Eq.}~\eqref{LLG}])$ \cite{thielePRL73}, upon substitution of ansatz \eqref{pa}. The key underlying physical assumption in this procedure is that the internal DW structure is dominated by the $A$ and $K_\ast$ terms in Eq.~\eqref{Hs}, such that it has a fixed width $\lambda_{\rm dw}\approx\sqrt{A/K_\ast}$, while the dynamics of slow variables $q_i$ are governed by the other terms in Eq.~\eqref{Hs}. Carrying out this program, we get (after a somewhat tedious but straightforward calculation) the following simple equations:
\begin{equation}
\dot{x}_{\rm dw}=-\frac{f_\phi+j_\ast\sin\phi_{\rm dw}}{(4+\alpha^2)S}\,,\,\,\,\dot{\phi}_{\rm dw}=-\frac{\alpha}{2\lambda_{\rm dw}}\dot{x}_{\rm dw}\,.
\label{eom}
\end{equation}
Here,
\begin{equation}
f_\phi\equiv-\frac{1}{L}\partial_{\phi_{\rm dw}}F=\frac{\eta_\ast\pi}{2}\sin\phi_{\rm dw}+h\lambda_{\rm dw}\pi\cos\phi_{\rm dw}
\label{ff}
\end{equation}
is the generalized force (per unit of DW length $L$) thermodynamically conjugate to the angle $\phi_{\rm dw}$. Since the domain wall is not pinned in the $x$ direction, the force $-\partial_{x_{\rm dw}}F$ conjugate to $x_{\rm dw}$ vanishes in our model. The energy dissipation $P\equiv-(\partial_{\phi_{\rm dw}}F)\dot{\phi}_{\rm dw}-(\partial_{x_{\rm dw}}F)\dot{x}_{\rm dw}$ associated with magnetic dynamics (in the absence of transport current $j_\ast$) is thus guaranteed to be positive in an out-of-equilibrium situation when $\alpha>0$. The spin-torque-driven DW dynamics in Eq.~\eqref{eom} reminds us of a dc Josephson effect ($\dot{Q}\propto\sin\varphi$). It is, in particular, noteworthy that the equilibrium and nonequilibrium spin torques add up, such that the latter can be formally absorbed into a redefinition of $\eta_\ast$: $\eta_\ast\to\eta_\ast+2j_\ast/\pi$. In the absence of the applied field, $h=0$, the dynamics would thus settle down at $\phi_{\rm dw}=0$ or $\pi$ (a N{\'e}el wall), for $\eta_\ast\lessgtr0$ (corresponding to the ordinary or $\pi$ Josephson junction, respectively). In the absence of spin torques but a finite field $h$ along the $y$ axis, the dynamics (that are overdamped as $\dot{\phi}_{\rm dw}\propto f_\phi$) would flow towards $\phi_{\rm dw}=\pm\pi/2$ (a Bloch wall), for $h\gtrless0$, which corresponds to the lowest magnetostatic energy.

Supplementing Eqs.~\eqref{eom} with the Onsager reciprocity principle \cite{landauBOOKv5}, dictates how the DW dynamics induce transport current along the chiral mode. (See Fig.~\ref{fig2}.) To infer this, consider a voltage $V_y$-induced current: $j_\ast=(g_QJ/ev)V_y$. From Eqs.~\eqref{eom} and \eqref{ff}, which describe how this voltage induces dynamics $(\dot{x}_{\rm dw},\dot{\phi}_{\rm dw})$, we recover their Onsager (time-reversed) counterpart in the charge sector:
\begin{align}
\dot{Q}&=\frac{g_QJ}{ev}\frac{\alpha\sin\phi_{\rm dw}}{2\lambda_{\rm dw}(4+\alpha^2)S}\partial_{\phi_{\rm dw}}F\nonumber\\
&\to-\frac{g_QJ}{ev}L\dot{\phi}_{\rm dw}\sin\phi_{\rm dw}=\frac{g_QJ}{ev}L\partial_tm_x(x_{\rm dw})\,,
\label{Qd}
\end{align}
where on the second line we  dropped the term that is diagonal in the charge sector and thus outside of the reciprocal reasoning \cite{Note4}. In the final equality of Eq.~\eqref{Qd}, we recognize exactly the Landauer-B{\"u}ttiker formula \eqref{IwLB} for the magnetization-dynamics-driven charge current. It is crucial to notice that the DW chirality flips under time reversal, as illustrated in Fig.~\ref{fig2}. The charge $Q$ in Eq.~\eqref{Qd} pumped by the DW dynamics in the top panel of Fig.~\ref{fig2} thus enters the reservoir that is opposite to the one where the voltage $V_y$ is applied, as must be since $\dot{\phi}_{\rm dw}$  certainly induces the current only downstream of the chiral mode. This proves internal consistency of our theory.

In summary, we developed a self-consistent hydrodynamic description of a magnetic DW bound with its parity-anomaly chiral electron mode. DW dynamics parametrized by slow variables $x_{\rm dw}$ and $\phi_{\rm dw}$ share similarities with ac/dc Josephson relations for charge and phase, respectively. In particular, the DW switches between two types of N{\'e}el walls (corresponding to $0$ and $\pi$ junctions) depending on the sign of the spin torque, and two types of Bloch walls depending on the sign of the applied field. Reciprocally, the chiral transport is pumped by the DW dynamics, in accord with fictitious gauge fields along the DW length. This coupled system provides a ballistic electron interconnect, which can be imprinted onto TI surfaces and dynamically controlled by magnetic fields, opening rich possibilities for ``magnetic lithography" of electronic nanostructures on TI surfaces.

This work was supported by the Alfred P. Sloan Foundation, DARPA, the NSF under Grant No. DMR-0840965 (Y.T.), and by the Swiss NSF (D.L.). Discussions with M.~Z. Hasan are gratefully acknowledged.

\begin{widetext}

\setcounter{equation}{0}

\section{Electromagnetic response of topological insulators}

A simple \textit{strong} topological insulator surface has a single Dirac cone of gapless electronic excitations \cite{hasanRMP10,*qiRMP11}. In the presence of a mirror-symmetry plane normal to the surface, and neglecting any warping effects, the effective low-energy Hamiltonian is then given by $H_0=v\mathbf{p}\cdot\mathbf{z}\times\boldsymbol{\hat{\sigma}}+\Delta\hat{\sigma}_z$. Here, $\mathbf{p}=(p_x,p_y)=-i\hbar\boldsymbol{\nabla}$ is the 2D canonical momentum in the plane of the surface, $\mathbf{z}$ normal unit vector, $\boldsymbol{\hat{\sigma}}=(\hat{\sigma}_x,\hat{\sigma}_y,\hat{\sigma}_z)$ vector of Pauli matrices for effective electron spin, $v$ Fermi velocity, and $\Delta$ (parity-breaking) mass term due to a time-reversal symmetry-breaking mechanism, which we take to be infinitesimal for the moment. The linear Dirac dispersion is valid up to a momentum cutoff $p_c\sim E_g/v$, where $E_g$ is the bulk spin-orbit gap of the TI. Minimally coupling electrons to the gauge \textit{3-potential} $A_\mu\equiv(\varphi,-\mathbf{A})$,
\begin{equation}
H_0\to v(\mathbf{p}-e\mathbf{A})\cdot\mathbf{z}\times\boldsymbol{\hat{\sigma}}+\Delta\hat{\sigma}_z+e\varphi\,,
\label{H}
\end{equation}
we recall the long-wavelength low-frequency charge response \cite{redlichPRL84,*jackiwPRD84}:
\begin{equation}
j^\mu={\rm sgn}(\Delta)(e^2/4h)\epsilon^{\mu\alpha\beta}F_{\alpha\beta}\,.
\label{j}
\end{equation}
Here, $\epsilon^{\mu\alpha\beta}$ is the antisymmetric Levi-Civita tensor with $\epsilon^{012}=1$, $F_{\alpha\beta}=\partial_\alpha A_\beta-\partial_\beta A_\alpha$ is the $3\times3$ field strength corresponding to the in-plane electric field $\mathbf{E}=-\partial_t\mathbf{A}-\boldsymbol{\nabla}\varphi$ and the normal magnetic field $B_z=\mathbf{z}\cdot\boldsymbol{\nabla}\times\mathbf{A}$. $j^0\equiv\sigma$ is the 2D surface charge density and $j^{1,2}$ are respectively the $x,y$ current-density components. The continuity equation $\partial_\mu j^\mu=0$ is, according to Eq.~\eqref{j}, guaranteed by the Faraday's law of induction (the Bianchi identity): $\mathbf{z}\cdot(\partial_t\mathbf{B}+\boldsymbol{\nabla}\times\mathbf{E})=0$. We can see this explicitly after rewriting Eq.~\eqref{j} in terms of the electromagnetic fields:
\begin{equation}
\sigma=-{\rm sgn}(\Delta)g_HB_z\,,\,\,\,\mathbf{j}={\rm sgn}(\Delta)g_H\mathbf{z}\times\mathbf{E}\,,
\label{rj}
\end{equation}
where $g_H\equiv e^2/2h=g_Q/2$ is the Hall conductance given by half of the conductance quantum $g_Q\equiv e^2/h$. Integrating out  electrons, would produce the Chern-Simons action for the electromagnetic field \cite{redlichPRL84}:
\begin{equation}
S_{\rm CS}=-(g_Q/4)\int dtd^2r\,{\rm sgn}(\Delta)\epsilon^{\mu\alpha\beta}A_\mu\partial_\alpha A_\beta\,.
\label{CS}
\end{equation}
One can easily verify that Eq.~\eqref{CS} is in one-to-one correspondence with the charge response \eqref{j}, as $j^\mu=-\delta_{A_\mu}S$.

The TI surface response \eqref{rj} is tied up with the axion electrodynamics of the TI bulk \cite{wilczekPRL87}: The full time-reversal-invariant electromagnetic Lagrangian in the presence of a TI bulk is given by $\mathcal{L}_{\rm EM}=\mathcal{L}_0-\theta\mathbf{E}\cdot\mathbf{B}$, where $\mathcal{L}_0$ is the ordinary (vacuum) Maxwell's Lagrangian and $\theta\mathbf{E}\cdot\mathbf{B}$ is the topological axion term with $\theta=sg_Q/2$, where $s={\rm sgn}(\Delta)$ inside the TI and 0 outside. The Maxwell's equations are then modified as follows: $\boldsymbol{\nabla}\cdot\mathbf{E}=\rho+\boldsymbol{\nabla}\theta\cdot\mathbf{B}$ and $\boldsymbol{\nabla}\times\mathbf{B}=\partial_t\mathbf{E}+\mathbf{J}-(\partial_t\theta+\boldsymbol{\nabla}\theta\times\mathbf{E})$, where $\rho$ and $\mathbf{J}$ are the ordinary external three-dimensional charge and current densities (entering $\mathcal{L}_0$). For a time-independent $\theta$, we thus find an additional 2D charge density $\sigma=-sg_QB_z/2$ and current density $\mathbf{j}=sg_Q\mathbf{z}\times\mathbf{E}/2$ at the TI surface, in accord with Eqs.~\eqref{rj}. Finally, it is useful to note that the sign $s$ ambiguity of $\theta$ inside of the TI when $\Delta\to0$ is relieved by the invariance of the full Lagrangian upon $\theta\to\theta+ng_Q$, where $n$ is an arbitrary integer \cite{wilczekPRL87}.

\section{Magnetic-insulator/topological-insulator exchange interaction}

Coupling the TI surface to a MI film with a dynamic magnetic (spin-density) texture $\mathbf{S}(\mathbf{r},t)=S\mathbf{m}(\mathbf{r},t)$ will induce time-reversal symmetry-breaking terms in the effective Hamiltonian $H=H_0+H'$, with the axially-symmetric local coupling given by the anisotropic exchange:
\begin{equation}
H'=J(m_x\hat{\sigma}_x+m_y\hat{\sigma}_y)+J_\perp m_z\hat{\sigma}_z\,.
\label{ex}
\end{equation}
Note that there is no \textit{a priori} reason to assume that $J=J_\perp$ as in Refs.~\cite{garatePRL10,*yokoyamaPRB10md,*nomuraPRB10,*uedaCM11}, based on symmetry considerations. The in-plane exchange induces an effective vector potential while the out-of-plane exchange opens a gap:
\begin{equation}
\mathbf{a}=(J/ev)\mathbf{m}\times\mathbf{z}~~~{\rm and}~~~\Delta=J_\perp m_z\,,
\end{equation}
respectively. It thus follows that a dynamic magnetic texture induces a charge response given (away from the $m_z=0$ regions) by \cite{nomuraPRB10}
\begin{equation}
\sigma={\rm sgn}(\Delta)\frac{g_QJ}{2ev}\boldsymbol{\nabla}\cdot\mathbf{m}\,,\,\,\,\mathbf{j}=-{\rm sgn}(\Delta)\frac{g_QJ}{2ev}\mathbf{z}\times\partial_t\mathbf{m}\times\mathbf{z}\,,
\label{jm}
\end{equation}
where $\boldsymbol{\nabla}$ should be understood as the 2D differential operator $(\partial_x,\partial_y)$ tangential to the TI surface. Note that there is a sign ambiguity of the Hall response \eqref{jm} when $m_z=0$ (and thus $\Delta$ changes sign), a point of central interest in this work.

\section{Chiral electron mode}

The TI is gapped away from the DW, while the region under the DW supports a chiral electron mode. It can be physically revealed by applying a uniform electric field $E$ along the $y$ axis (for a configuration sketched in Fig.~1 of the Main Text), which induces a Hall current $j_x=-{\rm sgn}(m_z)g_QE/2$ (supposing the magnetization is kept static and, to be specific, taking $J_\perp>0$ here). This, in turn, induces a nonequilibrium 1D electron charge density $\lambda$ along the DW, which is found by the continuity equation:
\begin{equation}
\partial_t\lambda=\int_-^+ dx\partial_t\rho=j_{x,-}-j_{x,+}=g_QE\,,
\label{dl}
\end{equation}
where $\pm$ designate domains at $x\gtrless x_{\rm dw}$, respectively. Here, we have invoked the Hall response \eqref{rj} away from the DW (which would break down in close vicinity of the point $m_z=0$). On the other hand, recall that a 1D chiral mode is excited by an electric field $E$ as follows: the acceleration $\dot{p}=eE$ induces a charge density increase $\partial_t\lambda=e\dot{p}/2\pi\hbar=g_QE$, in agreement with Eq.~\eqref{dl}.

In order to develop a microscopic understanding of the chiral mode and evaluate the relevant energy and length scales, we can explicitly solve the eigenvalue problem for the Hamiltonian
\begin{equation}
H=-i\hbar v(\hat{\sigma}_x\partial_y-\hat{\sigma}_y\partial_x)+\zeta x\hat{\sigma}_z\,,
\label{Hz}
\end{equation}
near the DW region where the gap $\Delta(x)=\zeta x$ ($\zeta\sim J_\perp/\lambda_{\rm dw}$) vanishes. The problem is fully analogous to calculating edge states of the integer quantum Hall effect \cite{halperinPRB82,*girvinCHA99}. As the Hamiltonian is translationally invariant in the $y$ direction, we look for energy $\varepsilon$ eigenvalues of the form $\hat{\psi}(x,y)=e^{ip_yy/\hbar}\hat{\varphi}(x)$, where the $x$-dependent part $\hat{\varphi}$ for the chiral mode can be easily found as \footnote{Solutions \eqref{cm} are easily generalized to an arbitrary gap profile, $\zeta x\to\Delta(x)$, in Eq.~\eqref{Hz} by replacing $\zeta x^2/2\to\int dx\Delta(x)$ in the exponent of $\hat{\varphi}_\pm$.}
\begin{equation}
\hat{\varphi}_\pm=e^{-(\zeta/2\hbar v)\hat{\sigma}_xx^2}\hat{u}_\pm\,,\,\,\,\varepsilon=\pm vp_y\,.
\label{cm}
\end{equation}
$\hat{u}_\pm$ here are the spinor eigenstates of $\hat{\sigma}_x$ with eigenvalues $\pm1$, respectively. Assuming, for concreteness, that $\zeta/v>0$, only the $\propto\hat{u}_+$ eigenstate with energy $\varepsilon=vp_y$ is physical, being localized as a Gaussian $\propto e^{-(\zeta/2\hbar v)x^2}$ near the origin. The characteristic width of the chiral mode is thus $\xi=\sqrt{\hbar v/\zeta}$: the sharper the DW the narrower the chiral mode. The corresponding gap between the chiral zero mode and the dispersive gapped states from the adjacent 2D regions can be estimated as
\begin{equation}
\delta_{\rm dw}\sim\xi\zeta=\sqrt{\hbar v\zeta}\,.
\label{ddw}
\end{equation}

The width of the chiral mode $\xi$ corresponds to a narrow region near $x_{\rm dw}$, which we assume to be much smaller than the DW width $\lambda_{\rm dw}$ in the ferromagnetic film (typically 10-100~nm \cite{tataraPRP08}). Accordingly, in addition to the 2D charge density captured by Eq.~\eqref{jm}, the chiral mode contributes 1D charge density
\begin{equation}
\partial_t\lambda(y)=g_Q\mathcal{E}_y=g_Q\mathbf{y}\cdot\left[\mathbf{E}+\frac{J}{ev}\mathbf{z}\times\partial_t\mathbf{m}(x_{\rm dw},y)\right]\,.
\label{dw}
\end{equation}
For a DW parametrized by Eqs.~(2) of the Main Text, $\partial_t\mathbf{m}(x_{\rm dw},y)\propto\mathbf{z}$, in the case of a sliding motion with $\dot{x}_{\rm dw}\neq0$ and $\dot{\phi}_{\rm dw}=0$, such that the DW dynamics-induced charge density \eqref{dw} vanishes. If $\dot{\phi}_{\rm dw}\neq0$, on the other hand, a finite electromotive force (EMF) $\propto\partial_t m_x$ is generally generated by the DW dynamics. In order to account, furthermore, for the 1D charge hydrodynamics along the DW, we would have to substitute $\partial_t$ on the LHS of Eq.~\eqref{dw} with the advective derivative: $\partial_t\to D_t\equiv\partial_t+v_\ast\partial_y$. $v_\ast$ here is the velocity associated with the chiral-mode dispersion $\varepsilon=v_\ast p_y$: $v_\ast=v(1+g_f)$, where $g_f$ parametrizes the (Luttinger-liquid) strength of the electron-electron forward scattering \cite{giamarchiBOOK04}.
Since the current carried by an excess density $\lambda$ is $I_{\rm dw}=v_\ast\lambda$, Eq.~\eqref{dw} gives
\begin{equation}
I_{\rm dw}(t)=g_Qv_\ast\int^tdt\left[E_y+\frac{J}{ev}\partial_tm_x(x_{\rm dw},t)\right]\,,
\label{jw}
\end{equation}
neglecting the aforementioned advective corrections $\propto v_\ast\partial_y$. The divergence of a dc current response to electric field $E_y$ reflects the absence of dissipation for a chiral mode, which is essentially insensitive to disorder. This changes profoundly, if we connect the chiral mode to Fermi-liquid reservoirs, as discussed in the Main Text.

\end{widetext}

\end{document}